\documentclass[aps,prd,preprintnumbers,nofootinbib,showpacs]{revtex4}
\usepackage{graphicx}
\usepackage{amsfonts}
\usepackage{amssymb,amsmath}
\usepackage{color}
\usepackage[colorlinks=true,linkcolor=blue,citecolor=blue]{hyperref}

\newcommand{\simgt}{\lower.5ex\hbox{$\; \buildrel > \over \sim \;$}}
\newcommand{\simlt}{\lower.5ex\hbox{$\; \buildrel < \over \sim \;$}}

\begin{document}
\begin{center}
{\Large \bf
What can we learn from higher multipole power spectra \\
of galaxy distribution in redshift space?}

\vskip .45in

{Tatsuro Kanemaru${}^1$, 
Chiaki Hikage${}^2$, Gert H\"utsi${}^3$, 
Ayumu Terukina${}^1$, Kazuhiro Yamamoto${}^{1}$
}

\vskip .45in

{\em
${}^1$Department of Physical Sciences, Hiroshima University, 
Kagamiyama 1-3-1, \\
Higashi-hiroshima, 
739-8526, Japan
\\
${}^2$Kobayashi-Masukawa Institute, Nagoya University, 
Nagoya 464-8602, Japan
\\
${}^3$Tartu Observatory, T\~oravere 61602, Estonia
}

\end{center}

\begin{abstract}
We investigate a potential of the higher multipole power spectra of the galaxy 
distribution in redshift space as a cosmological probe on halo scales. 
Based on the fact that a halo model explains well the multipole power spectra 
of the luminous red galaxy (LRG) sample in the Sloan Digital Sky Survey (SDSS), 
we focus our investigation on the random motions of the satellite LRGs that 
determine the higher multipole spectra at large wavenumbers. 
We show that our theoretical model fits the higher multipole spectra at 
large wavenumbers from N-body numerical simulations and we apply these results for testing the gravity theory and the velocity structure
of galaxies on the halo scales.
In this analysis, we use the multipole spectra $P_4(k)$ and $P_6(k)$
on the small scales of the range of wavenumber
$0.3 \leq k/[h{\rm Mpc}^{-1}]\leq 0.6$, which is in contrast 
to the usual method of testing gravity by targeting the linear growth rate on very 
large scales. 
We demonstrate that our method could be useful for testing gravity on the halo scales. 
\end{abstract}

\maketitle

\setcounter{page}{1}
\label{sec:intro}
Redshift survey of galaxies is a promising way for exploring the nature the dark energy 
and testing gravity on the cosmological scales. 
Recent results of the baryon oscillation spectroscopic survey (BOSS) date release (DR) 
11 of the Sloan Digital Sky Survey (SDSS) III have demonstrated the usefulness of 
redshift surveys \cite{Anderson,Percival}.
A possible tension in the cosmological parameters between the results by the Planck 
satellite and the BOSS is reported \cite{Samushia,Chuang,Sanchez,Beutler}, which attracts 
interests of researchers.
An interesting question that arises is whether this tension could be resolved in models
where the gravity gets modified from its usual general relativistic form.

The redshift-space distortion plays an important role in testing gravity \cite{Guzzo,YSH}, 
which reflects the information of velocity of galaxies. 
One of the target of redshift surveys is a measurement of the redshift-space 
distortions in the linear regime of the density perturbations 
\cite{Hikage2014}, which provides us with a chance of testing gravity through the linear 
growth rate.  
On the other hand, the Finger-of-God (FoG) effect is the redshift-space distortion in the 
nonlinear regime of density perturbations reflecting the random motion of galaxies.
The primary purpose of the present paper is to investigate an effective method 
to evaluate the random velocity of galaxies in halos, which might 
provide us with a unique chance of testing gravity on halo scales. 
This can be achieved by precisely modeling the FoG effect on the basis 
of the halo model.  

In order to quantify the redshift-space distortions, the multipole power spectrum
is useful, which is defined as a multipole coefficient of the multipole expansion 
of the anisotropic power spectrum (e.g., \cite{Yamamoto2006,YSH,Beutler}). 
Recently, the authors of Ref.~\cite{HY} found that a halo model describes well 
small-scale behavior of the higher multipole power spectra of the luminous red 
galaxy (LRG) sample of SDSS DR7.
Based on this new finding, we consider a potential of measuring the velocity of 
satellite galaxies in halos and testing the gravity theory on the halo scales with 
the multipole power spectrum. 
The key of the method is the random motion of the satellite galaxies and their 
1-dimensional velocity dispersion in a halo with mass $M$, for which we adopt a 
simple formula, 
\begin{eqnarray}
&&\sigma^2_{\rm v}(M)=\beta{GM\over 2r_{\rm vir}},
\label{sigmav2}
\end{eqnarray}
where $\beta$ is a constant parameter, $G$ is the Newton's universal 
gravitational constant, and $r_{\rm vir}$ is the virial radius defined 
by $r_{\rm vir}=(3M/4\pi\bar{\rho}_{\rm m}(z)
\Delta_{\rm vir}(z))^{1/3}$, where $\bar\rho_{\rm m}(z)$ is the mean 
matter density and $\Delta_{\rm vir}(z)$ is the density 
contrast of a halo, respectively, at the redshift $z$. 
We adopt $\Delta_{\rm vir}=265$ at $z=0.3$, for the sample 
corresponding to our LRG mock samples. 
We carefully check this velocity dispersion relation using 
the numerical simulations, as well as the validity of the theoretical model 
for the higher multipole power spectra.  
This theoretical model is compared with the SDSS LRG sample, and we put a 
useful constraint on the velocity dispersion and the gravitational constant 
on the halo scales. 

\label{sec:fm}
We here briefly review the multipole spectrum in a 
halo model according to Ref.~\cite{HY,Hikage2014}. 
Following the general prescription of the halo 
approach~\cite{Seljak2001,White2001,CooraySheth2002}, 
we write the anisotropic power spectrum in the redshift-space 
consisting of the 1-halo and 2-halo terms,
$
P_{\rm LRG}(k,\mu)=P^{\rm 1h}(k,\mu)+P^{\rm 2h}(k,\mu).
$ 
We here consider the model which consists of the central galaxies and
the satellite galaxies. We adopt the following expression
(\ref{eq:pk_1h}) for the one halo term,
\begin{eqnarray}
&&P^{\rm 1h}(k,\mu)=\frac{\displaystyle 1}{\displaystyle \bar{n}^2}\int\!dM~
\frac{\displaystyle dn}{\displaystyle dM}
\Bigl[2\langle N_{\rm sat}\rangle
\tilde u_{\rm NFW}(k,M) e^{-(\sigma_c^2+\sigma_s^2) k^2\mu^2/2a^2H^2}
\nonumber\\
&&\hspace{5cm}+\langle N_{\rm sat}\rangle^2
\tilde u^2_{\rm NFW}(k,M) e^{-\sigma_s^2 k^2\mu^2/a^2H^2}
\Bigr],
\label{eq:pk_1h}
\end{eqnarray}
where we adopt the halo mass function $dn/dM$ given by
\cite{ShethTormen1999}, $\bar{n}$ is the mean number density of
LRGs given by $\bar{n}=\int dM (dn/dM) N_{\rm HOD}(M)$, $N_{\rm
  HOD}(M)=\langle N_{\rm cen}\rangle +\langle N_{\rm sat}\rangle$ is 
the halo occupation distribution (HOD), 
for which we adopt the following 
form
\cite{Zheng2005}, 
\begin{eqnarray}
&&\langle N_{\rm cen}\rangle =\frac{1}{2}\left[1+{\rm erf}\left(\frac{\log_{10}(M)-\log_{10}
(M_{\rm min})}{\sigma_{\log M}}\right)\right], \\
&&\langle N_{\rm sat}\rangle =
  \langle N_{\rm cen}\rangle \left(\frac{M-M_{\rm cut}}{M_1}\right)^{\alpha},
\label{eq:HOD}
\end{eqnarray}
with the error function ${\rm erf}(x)$, and 
$\sigma^2_c(M)$ 
and $\sigma^2_s(M)$ are the velocity dispersion of the central LRGs
and the satellite LRGs, respectively. 
Table 1 lists the HOD parameters matching the SDSS DR7 LRG catalog 
in Ref.~\cite{Reid2009a}. 
We assume that the distribution of the satellite galaxies follows
the NFW profile \cite{NFW1996} and $\tilde{u}_{\rm NFW}(k)$ denotes the
Fourier transform of truncated NFW profile \cite{Scoccimarro2001}.
Results of Ref.~\cite{Guo} support this assumption. 
%
We may assume that central LRGs reside near the halo center, thus their 
velocity difference relative to the host halo should be small 
(cf.~\cite{Hikage2012}). 
On the other hand, satellite LRGs are off-centered and their random velocity 
should be the main source of the FoG effect. 
Here we assume 
\begin{eqnarray}
&&\sigma^2_c(M)=\alpha_c^2 \sigma^2_{\rm v}(M),
\\
&&\sigma^2_s(M)=\alpha_s^2 \sigma^2_{\rm v}(M),
\end{eqnarray}
where $\alpha_c$ and $\alpha_s$ are the constant parameters. 

In the previous paper \cite{HY}, the 2-halo term was modeled with an 
analytic fitting formula from N-body simulations. 
However, in the present paper, we adopt a very simple treatment
for the two halo term of the higher multipole power spectrum,
because it is not trivial to construct a precise analytic model in 
redshift space which is applicable even at large wavenumbers. 
Using the mock catalogs corresponding the LRG sample 
is an alternative way to incorporate precise theoretical predictions
for the two halo term. For this modeling, we adopt
the results in the previous paper \cite{Hikage2014}, which has
constructed mock catalogs, corresponding to the SDSS LRG sample, 
and has investigated the behavior of the multipole spectra. 
In the present paper, we use the following modeling for 
$P^{2h}_4(k)$ and $P^{2h}_6(k)$. 
The results in \cite{Hikage2014} demonstrate
that the contribution from the two halo term to $P_6(k)$ is negligible, 
i.e., $P^{2h}_6(k)\simeq 0$, and $P_4^{2h}(k)$ is simply expressed 
as $kP^{2h}_4(k)\simeq 15 [h{\rm Mpc}^{-1}]^2$, which we also adopt here. 
Because the contribution of the two halo term to $P_2(k)$ is rather 
large compared to that of $P_4(k)$ and $P_6(k)$ \cite{HY}, it
is not included in our analysis. 

\begin{table*}[h]
\begin{center}
\begin{tabular}{lc}
\hline
\hline
~ & Simulation/LRG \\
\hline
$M_{\rm min}$ &  $5.7\times 10^{13}M_\odot/h$\\
$\sigma_{\log M}$  & 0.7 \\
$M_{\rm cut}$ & $3.5\times 10^{13}M_\odot/h$ \\
$M_1$ & $3.5\times 10^{14}M_\odot/h$ \\
$\alpha$ & 1 \\
\hline
\end{tabular}
\caption{
HOD parameters of the LRG sample \cite{Reid2009a}.}
\end{center}
\label{tab:lrgHOD}
\end{table*}

We first demonstrate the validity of our theoretical model by comparing 
with the results of N-body simulations. 
The simulations assume the spatially flat cold dark matter model with 
a cosmological constant, adopting $\Omega_0=0.273$ and $\sigma_8=0.82$.
We run 10 realizations of N-body simulations using Gadget-2 code \citep{Springel05} with Gaussian 
initial condition. 
Each simulation has a side length of 600$h^{-1}$Mpc and the particle
  number of $800^3$ (each particle mass is $2.8\times
  10^{10}h^{-1}M_\odot$). We use $z=0.3$ snapshots and identify halos by 
Friends-of-Friends algorithm with a linking length of 0.2.
Mock catalogs are constructed so that the bias and the HOD match
the SDSS DR7 LRG catalog in Ref.~\cite{Reid2009a}. 
The position of a central LRG is given by the potential minimum of the host halo and 
the velocity is given as the averaged velocity of all particles within the halos.
We substitute randomly-picked up dark matter particles for satellite LRGs.
In this analysis we constructed mock samples both with and without 
including the fiber collision effect \cite{Blanton03}.
We first make uncollided samples by removing one of the adjacent subhalos within 55 arcsec at $z=0.3$
and randomly return a part of removed subhalos at 10 percent probability for the overlapped area of tiling where 
both spectrum of collided pairs can be measured.
In our simulation, the central LRGs locate near the halo center, and their velocity is negligible. 
We assume no velocity bias for satellites. Thus, 
our mock catalogs should be understood as $\alpha_c=0$ and $\alpha_s=1$. 

Using the mock catalogs, we show the validity of our expression~(\ref{sigmav2})
for the velocity dispersion of satellite galaxies.
The velocity dispersion of satellite galaxies in a halo has not been well 
understood, though there are a few works that investigate the velocity 
dispersion of LRGs \cite{Hikage2012b,Masaki}. Recently, Guo et al. have 
studied the velocity bias of galaxies in the SDSS III CMASS sample
in the context of a halo model \cite{GuoVD}. Their results have implications
for our results, as will be discussed below. 

\begin{figure}[t]
\begin{center}
\vspace{.0cm}
    \hspace{0mm}\scalebox{.45}{\includegraphics{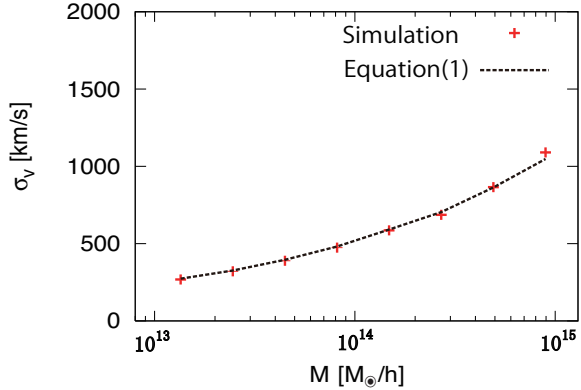}}
\vspace{0.cm}
\caption{1-dimensional velocity dispersion $\sigma_{\rm v}(M)$ as a function of halo mass.
The crosses are from N-body simulation, while the curve is Eq.~(\ref{sigmav2}) with $\beta=1$.
  \label{fig:sigmav}}
\end{center}
\end{figure}

Figure \ref{fig:sigmav} compares the velocity dispersion $\sigma_{\rm v}^2(M)$
of satellites as a function of the host halo's mass $M$. Here
the cross symbols show the results of the N-body numerical simulation, 
while the curve shows $(GM/2r_{\rm vir})^{1/2}$, i.e., Eq.~(\ref{sigmav2})
with $\beta=1$. 
This suggests that Eq.~(\ref{sigmav2}) with $\beta=1$ reproduces well the relation 
between the velocity dispersion of satellite and the halo mass of our N-body 
simulations. 

The effect of the fiber collision, which misses galaxies located closely to
each other, could be crucial in the analysis of the redshift-space 
clustering on small scales \cite{Guofiber}. 
The fiber collision dominantly occurs for pairs in the same halo.   
In the previous work \cite{HY}, the effect of the fiber collision
is included by a multiplying factor reducing the satellite fraction. 
In the present paper, we adopt a similar prescription, for simplicity.
Instead of introducing the satellite fraction, we float the HOD parameter 
$M_1$, which changes the satellite fraction, as a fitting parameter in our 
MCMC analysis. 

\begin{table}[h]
\begin{center}
\begin{tabular}{ccccc}
\hline
\hline
~~~~~~~~ & (A)~Mock with F.C. & (B)~Mock without F.C. & ~~~~(C)~LRG~~&  ~~(D)~LRG with BLRG~~\\
\hline
$\beta$ & ~$1.17^{+0.56}_{-0.40}~\bigl(0.96^{+0.43}_{-0.30}\bigr)$  
& $0.97^{+0.30}_{-0.24}~\bigl(0.83^{+0.23}_{-0.20}\bigr)$ 
& $1.70^{+0.83}_{-0.55}~\bigl(1.79^{+0.83}_{-0.59}\bigr)$ 
& $1.35^{+0.68}_{-0.45}~\bigl(1.38^{+0.65}_{-0.47}\bigr)$ 
\\
$~M_{\rm 1}[10^{14}M_\odot/h]$ & ~$6.5^{+0.9}_{-1.0}~\bigl(6.1^{+1.0}_{-1.1}\bigr)$  
&  $4.1^{+0.4}_{-0.4}~\bigl(4.0^{+0.6}_{-0.6}\bigr)$ 
&  $4.0^{+0.4}_{-0.4}~\bigl(4.0^{+0.4}_{-0.4}\bigr)$ 
&  $4.0^{+0.4}_{-0.5}~\bigl(4.0^{+0.4}_{-0.5}\bigr)$ 
\\
{\small \rm satellite ~fraction(\%)}&~ $3.8^{+0.7}_{-0.5}~\bigl(4.1^{+0.8}_{-0.5}\bigr)$  
&  $5.9^{+0.6}_{-0.5} ~\bigl(6.2^{+0.7}_{-0.6}\bigr)$ 
&  $6.3^{+0.5}_{-0.4} ~\bigl(6.3^{+0.5}_{-0.4}\bigr)$ 
&  $6.4^{+0.5}_{-0.4} ~\bigl(6.4^{+0.5}_{-0.4}\bigr)$ 
\\
$\chi^2$ &~ $16 ~(58)$  
&   $18 ~(59)$ 
&  $47 ~(56)$ 
&  $6.4 ~(7.3)$ 
\\
d.o.f. &~ $10 ~(12)$  
&   $10 ~(12)$ 
&  $60 ~(80)$ 
&  $60 ~(80)$ 
\\
\hline\hline
\end{tabular}
\caption{
Results of our MCMC analysis with floating the $2$ parameters $\beta$ and $M_1$, 
where we fixed $\alpha_c=0$ and $\alpha_s=1$ and the other cosmological 
parameters. 
The best fitting values with one sigma statistical errors, the satellite fraction 
$\int dM (dn/dM)\langle N_{\rm sat}\rangle/\bar n$, 
and the chi-squared 
along with the number of d.o.f are presented when fitted with (A) the simulation with the 
fiber collision (F.C.), (B) the simulation without the fiber collision,  
(C) LRG sample and (D) LRG sample with the two halo term modeled using the BLRG 
sample, from the left to the right column, respectively.  
The results are obtained using the data in the range 
of the wavenumbers $0.3\leq k/[h{\rm Mpc}^{-1}]\leq 0.6$, and 
the values in the parenthesis are same but using the data 
in the range of the wavenumbers $0.2\leq k/[h{\rm Mpc}^{-1}]\leq 0.6$.
}
\end{center}
\label{tab:simulationnop0}
\end{table}

\begin{figure}
  \begin{center}
  \includegraphics[width=150mm]{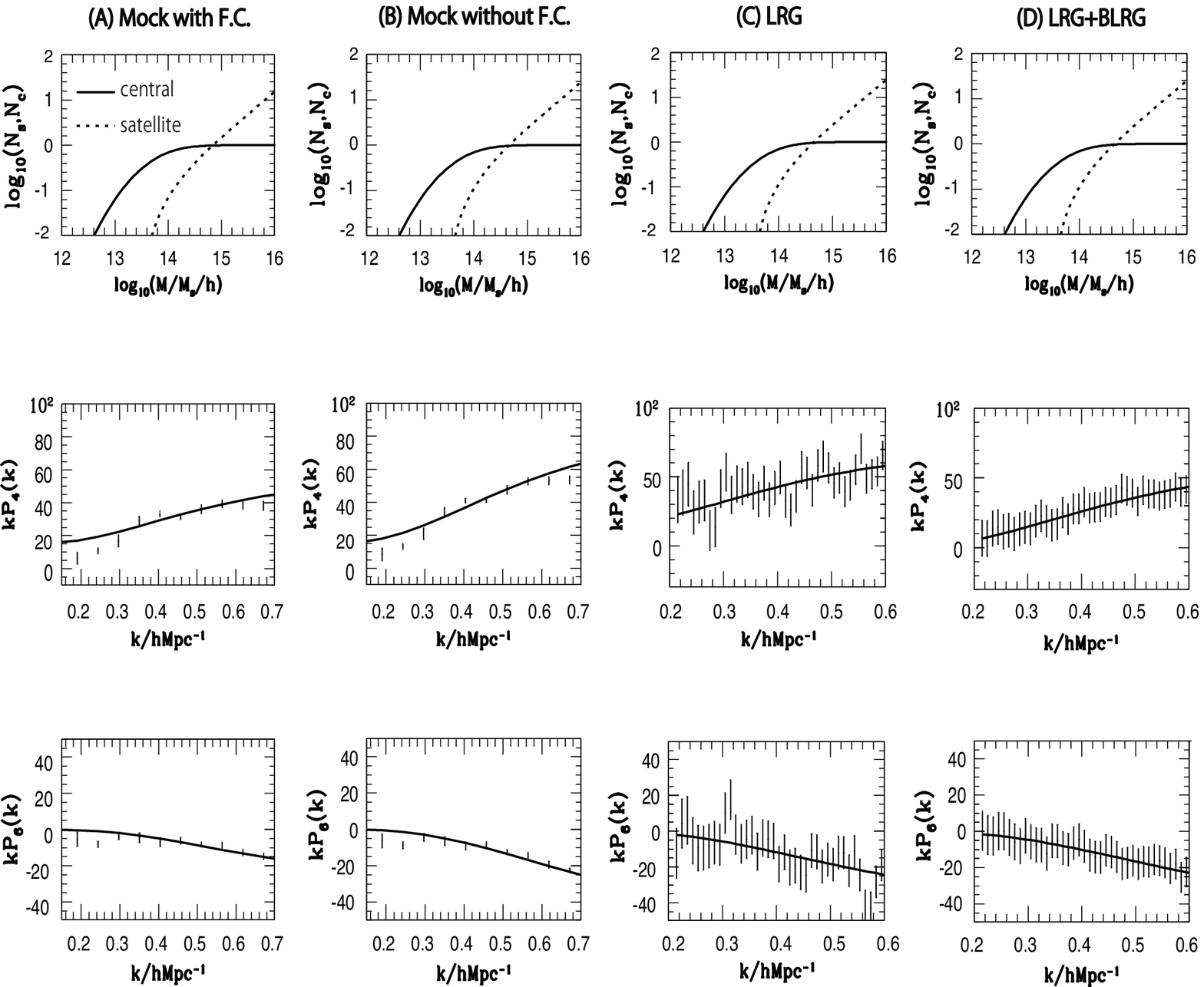}
  \end{center}
  \caption{
Top panels show the HOD $N_s(M)$ (dotted lines) and  $N_c(M)$ (solid lines), 
middle panels $kP_4(k)$, and bottom panels $kP_6(k)$.
In each panel, the curves correspond to the best fitting models.
Each column shows the results of (A) mock with fiber collision, (B)
mock without fiber collision, (C) LRG sample, and (D) LRG sample subtracted
the two halo term modeled using the BLRG sample, from left to right, 
respectively. 
}
  \label{fig:234}
\end{figure}

We compare the results of the numerical simulations and the theoretical 
multipole power spectra, with floating the two parameters $\beta$ and $M_1$.
Especially, we here fixed $\alpha_c=0$ and $\alpha_s=1$ taking the 
consistency with our numerical simulations. 
Note that the HOD parameters other than $M_1$ are fixed. 
In the MCMC analysis we only use $P_4(k)$ and $P_6(k)$ in the range of 
wavenumbers $0.3\leq k/[h{\rm Mpc}^{-1}]\leq0.6$ in order to 
reduce the influences from the uncertain contribution of the two halo term. 
%
%
Table II summarizes our results, 
where the best fitting values with one sigma statistical errors 
are presented for (A) simulation with the fiber collision (F.C.), 
(B) simulation without the fiber collision, and (C) LRG sample, (D) LRG 
sample with the two halo term modeled with the brightest LRG (BLRG) sample \cite{HY}, 
from the left to the right column, respectively. 
The chi-squared and the degrees of freedom are also shown. 
In this table, the values within 
parentheses are the results for the data in the range of  
$0.2\leq k/[h{\rm Mpc}^{-1}]\leq 0.6$. 
The left hand two columns of Figure \ref{fig:234} show the best 
fit curves of the HOD and the multipole power spectra for the simulations (A) 
and (B). These results demonstrate that our theoretical model reproduces 
those of the numerical simulations. 



We next apply our method to the multipole power spectra 
measured with the SDSS DR7 \cite{Abazajian}. 
The DR7 LRG sample is selected to cover a redshift range, 
$0.16 < z < 0.36$, only in the northern cap in order to reduce 
systematic uncertainties and to match the analysis in Ref.~\cite{Reid2009a}. 
Thus, the sky coverage is limited to $7189 ~{\rm deg.}^2$ and the total 
number of LRGs is $61\,899$. We adopt the same method for the measurement 
as that in Refs.~\cite{Oka,HY,Yamamoto2010}, but with the 
fiducial cosmological background for the distant-redshift relation 
of the spatially flat $\Lambda $CDM cosmology with $\Omega_m = 0.3$.
The right hand two columns in Table II list the results of the MCMC 
analysis with the LRG multipole spectra, whose difference comes from 
the modeling for the two halo term. 
The right two columns of Figure \ref{fig:234} show the best 
fitting curve and the data. 

The results of MCMC analysis with the SDSS LRG sample can be used for testing 
the gravitational constant on the halo scales. This is 
because the velocity dispersion in a modified gravity models
could be written as $\sigma^2_{\rm v}(M)={G_{\rm eff}M/ 2r_{\rm vir}}$,
where $G_{\rm eff}$ is an effective gravitational constant. 
Regarding $G_{\rm eff}=\beta G$, we may put a constraint on the effective 
gravitational constant from the SDSS LRG sample on the halo scales, 
$\beta =1.70^{+0.83}_{-0.55}$ from the column (C) in Table II, in which 
we adopted the same modeling for the two halo term as that of the mock catalogs.
This value is rather larger than the prediction of the numerical simulations, 
although the error is not small.

Though the contribution of the two halo term to $P_4$ and $P_6$ 
is rather small compared with the one halo term, but it might
be influential to our results. As a check of our results, we 
model the contribution of the two halo term using the BLRG 
sample \cite{HY}. 
Because the BLRG catalog roughly corresponds to the central galaxies
catalog, then we may model the two halo term by computing 
the multipole spectrum of the BLRG catalog. The column (D) of Table II 
represents the results $\beta =1.35^{+0.68}_{-0.45}$. 
Compared with the case of the modeling for the two halo term
from the numerical simulation, the value of $\beta$ becomes small 
and $\beta=1$ is in the one-sigma error of the results. 

Let us discuss the reason why higher value of $\beta$ is obtained 
from the analysis of the LRG sample. It could be a smoking gun of a modified gravity. 
For example, an $f(R)$ gravity model has an effective gravitation constant 
$G_{\rm eff}=4G/3$, as long as the chameleon mechanism does not work. 
However, we should discuss possible systematics that may lead to a larger 
value of $\beta$. Because the satellite fraction is small, being around $6$\%
of the total LRGs, the first term dominates the right hand side of 
Eq.~(\ref{eq:pk_1h}). Then, taking the degeneracy in the 
central and the satellite galaxy velocities, we should understand that the constraint is 
\begin{eqnarray}
  {\sigma_c^2+\sigma_s^2 \over \sigma_{\rm v}^2}=\beta(\alpha_c^2+\alpha_s^2)=1.35^{+0.68}_{-0.45},
\end{eqnarray}
in the case (D) when we use the BLRG sample for modeling the two halo term. 
The results might be explained by a larger velocity dispersion of the central galaxy
in the multiple system. Recently, Guo et al. have reported the velocity bias of galaxies in the 
SDSS CMASS samples \cite{GuoVD}. The sample is different from ours, but they report that 
$\alpha_c\sim0.3$. However, this value $\alpha_c\sim0.3$ is rather small to explain our results, 
and $\alpha_c\sim0.6$ is required within the general relativity $\beta=1$. 
Other possible systematics is the modeling of the two halo term in $P_\ell(k)$, 
as we obtained somewhat different values between (C) and (D) in Table II. 
More sophisticated simulations based on subhalo catalogs from 
N-body simulations could be necessary and useful. 
We checked that our treatment of the fiber collision effect works within 
the error. However, this effect could be more complicated and 
a more careful modeling of the fiber collision might be necessary. 

In summary, we have investigated the potential of the higher multipole 
power spectra of the galaxy distribution in redshift-space.
This method is based on the recent finding that a halo model accounts well for 
the behavior of the multipole power spectrum of LRGs on small scales.
Our method uses the data of the spectrum on small scales 
$0.3\leq k/[h{\rm Mpc}^{-1}]\leq 0.6$. 
This is quite in contrast to the usual method of testing gravity by measuring 
the linear growth rate on very large scales. 
Our method is based on the fact that one halo term makes a dominant 
contribution to the higher multipole power spectra at large wavenumbers, 
which reflects the random motions of the satellite galaxies. 
We carefully investigated the relation between the velocity dispersion 
of the random motions of satellite galaxies and the host halo mass 
on the basis of the mock catalogs from N-body simulations.
The validity of our theoretical model for the higher multipole power 
spectrum is tested using the results of the mock catalogs. 
By confronting our theoretical model and the observed multipole spectra 
of the SDSS LRG samples, we obtained a value for an effective gravitational 
constant somewhat larger than that predicted by the numerical simulations. 
This could be a smoking gun of the modified gravity. 
However, we might need to check our theoretical model for 
the two halo term and the fiber collision effect more carefully.

\vspace{2mm}
K. Y. thanks the workshop, APC-YITP collaboration: 
Mini-Workshop on Gravitation and Cosmology, which was held 
at YITP Kyoto universe for a useful chance for discussions 
on the topic of the present paper. This work is supported 
by a research support program of Hiroshima University. 
The research by C.H. is supported 
in part by Grant-in-Aid for Scientific researcher of Japanese 
Ministry of Education, Culture, Sports, Science, and Technology 
(No.~24740160).


\end{document}